\documentclass[12pt]{article}
\usepackage{graphicx}
\usepackage{epstopdf}
\usepackage{psfrag}                  
\usepackage{url}                        
\usepackage{times}                    
\usepackage{makeidx}               
\usepackage{multicol}                
\usepackage[bottom]{footmisc} 
\usepackage{fancyhdr}
\RequirePackage{subfigure}
\newcommand{\be}{\begin{equation}}
\newcommand{\ee}{\end{equation}}
\newcommand{\ba}{\begin{eqnarray}}
\newcommand{\ea}{\end{eqnarray}}

\newcommand{\bt}{\beta}
\newcommand{\vp}{\varphi}

\newcommand{\al}{\alpha}

\newcommand{\prt}{\partial}

\begin{document}

\begin{center}
{\Large{\bf Stable States of Biological Organisms} \\[5mm]
V.I. Yukalov$^{1,2}$, D. Sornette$^1$, E.P. Yukalova$^{1,3}$, 
J.-Y. Henry$^4$, \\
and J.P. Cobb$^5$} \\ [3mm]
{\it
$^1$Department of Management, Technology and Economics, \\
Swiss Federal Institute of Technology, CH-8032 Z\"urich, Switzerland\\ [2mm]

$^2$Bogolubov Laboratory of Theoretical Physics, \\
Joint Institute for Nuclear Research, Dubna 141980, Russia \\ [2mm]

$^3$Department of Computational Physics, Laboratory of Information
Technologies, \\
Joint Institute for Nuclear Research, Dubna 141980, Russia \\ [2mm]

$^4$Institut de M\'edecine et Sciences Humaines, \\
10 route de Bremblens, CH-1026 Echandens, Switzerland \\ [2mm]

$^5$ Cellular Injury and Adaptation Laboratory, \\
Washington University, St. Louis, USA}

\end{center}

\vskip 0.5cm

\begin{abstract}

A novel model of biological organisms is advanced, treating an organism 
as a self-consistent system subject to a pathogen flux. The principal 
novelty of the model is that it describes not some parts, but a biological
organism as a whole. The organism is modeled by a five-dimensional 
dynamical system. The organism homeostasis is described by the evolution 
equations for five interacting components: healthy cells, ill cells, 
innate immune cells, specific immune cells, and pathogens. The stability 
analysis demonstrates that, in a wide domain of the parameter space, the 
system exhibits robust structural stability. There always exist four 
stable stationary solutions characterizing four qualitatively differing 
states of the organism: alive state, boundary state, critical state, and 
dead state.

\end{abstract}

{\parindent=0pt
{\bf PACS}: 82.39.Rt; 87.10.Ca; 87.10.Ed; 87.18.-h

\vskip 1cm


\vskip 1cm

{\bf Corresponding author}:

\vskip 2mm

Prof. V.I. Yukalov

\vskip 1mm

{\bf E-mail}: yukalov@theor.jinr.ru }

\newpage

\section{Introduction}

Biological organisms are among the most complex systems, requiring for
their mathematical description rather elaborate equations [1--3]. Such
equations are, as a rule, nonlinear, because of which they can possess
different solutions, depending on the system parameters. The solutions 
of nonlinear equations are known to be very sensitive to the values of 
the control parameters. Sometimes, a slight variation of a parameter 
results in a discontinuous change of the system behavior, which is 
manifested in the qualitative change of its phase portrait. Such abrupt 
transformations of the phase diagram demonstrate the structural instability 
of the related dynamical system and are also called catastrophes [4]. In 
many cases, these catastrophes describe a real morphogenesis occurring in 
complex systems under the variation of the parameters. However, structural 
instability may also happen as an artifact, simply because a real complex 
system has not been correctly modelled by a dynamical system. It is 
therefore extremely important to carefully take into account the basic 
features of the real system, when modelling it by mathematical equations. 
Even a small term can essentially influence the behavior of a nonlinear 
dynamical system [5]. This especially concerns the terms reflecting 
fundamental symmetries of the complex system.

In the present paper, we consider such a very complicated biological system
as an organism consisting of five components, healthy cells, ill cells, two
types of immune cells, innate and specific, and pathogens. Our aim is twofold.
First, we formulate the dynamical system describing the homeostasis of an 
organism as a whole, paying attention to the necessity that the 
action-counteraction exchange symmetry be preserved. The principal point in 
our description is the treatment not of some parts of an organism, but the 
study of the latter as a self-consistent system. Second, we perform the 
stability analysis and demonstrate that the system exhibits a remarkable 
structural stability: in a very wide range of parameters, there always exist 
only four stable stationary solutions characterizing four organism states: 
alive state, boundary state, critical state, and dead state. Such a 
structural stability found in the proposed dynamical system provides an 
important validation step for the model [6], as it captures one of the most 
important characteristics of biological organisms, that of adapting robustly to 
different conditions.

\section{Construction of evolution equations}

Let us consider biological species, enumerated with the index $i=1,2,
\ldots$, the number of agents of the $i$-th type being $N_i$. Evolution 
equations can be represented either by differential or difference 
equations. Keeping in mind a very large number of interacting agents and 
the interaction times short as compared to the observation time, we employ 
here a picture with continuous time. The general structure of the dynamical 
system, describing the evolution of the species, can be represented as a 
set of differential equations
\be
\label{eq1}
\frac{dN_i}{dt} = R_i N_i \; + \; \sum_i R_{ij} N_i N_j \; +
\; F_i \; ,
\ee
where $t$ is time, $R_i$ is a life rate, $R_{ij}$ is  an interaction
intensity, and $F_i$ is an influx. The quantities $R_i$ and $R_{ij}$
are treated as parameters. The influx $F_i$ describes an input that is
external with respect to the $i$-type species. That is, $F_i$ may include
a flux that is external for the organism as a whole and also it may
describe a flux from other species. Generally, $F_i$ can be a function
of $N_j$, with $j\neq i$, such that
\be
\label{eq2}
\frac{\prt F_i}{\prt N_i} = 0 \; .
\ee

To be self-consistent, the set of equations (\ref{eq1}) has to satisfy
two major requirements. First of all, the various processes included in
the consideration must be of the same order of nonlinearity. This means
that, since the interaction term in Eq. (\ref{eq1}) is of second order,
the fluxes $F_i$ have also to be not higher than of second order with
respect to $N_j$, with $j\neq i$.

Another important requirement is the existence of the {\it
action-counteraction symmetry}. This means that, if in Eqs. (\ref{eq1})
there occurs a term $R_{ij}N_iN_j$ then, these equations have to contain
the exchange term $R_{ji}N_jN_i$. That is, if there is an action, there
should exist a counteraction, which can be schematically formulated as
the symmetry
\be
\label{eq3}
R_{ij} N_i N_j \; \leftrightarrow \; R_{ji} N_j N_ i \; .
\ee
Note that $R_{ij} \neq R_{ji}$ in general.

To specify a biological system, we consider an organism consisting of $N_1$
healthy cells, $N_2$ ill cells, $N_3$ innate immune cells, $N_4$ specific
immune cells, and $N_5$ pathogens. The organism homeostasis is characterized
by the processes and reactions between the cells and pathogens. A detailed
medical description of these processes and reactions is given in Ref. [7].
Here, we shall list them only in brief, specifying the parameters $R_j$ and
$R_{ij}$, with taking account of the typical signs of these parameters.

\vskip 2mm
(i) {\it Healthy cells} are characterized by a natural reproduction rate
$R_1\equiv A_1$. Since the body volume is limited, the carrying capacity
limitation is governed by $R_{11}\equiv-A_{11}$. Ill cells do not interact
directly with healthy cells, which implies that $R_{12}=0$. Healthy cells
can be occasionally attacked by innate as well as specific immune cells,
hence $R_{13}\equiv-A_{13}$ and $R_{14}\equiv-A_{14}$, which causes
autoimmune diseases. Pathogens infect healthy cells, thus $R_{15}\equiv
-A_{15}$. We treat an organism as a self-organizing system, which means
that healthy cells can be reproduced inside the organism, but are not
supplied from exterior, that is, $F_1=0$.

\vskip 2mm
(ii) {\it Ill cells} have either a natural death rate $R_2\equiv-A_2$,
when $A_2>0$, or can exhibit an unnatural proliferation, when $A_2<0$,
which happens under certain diseases, such as cancer. Healthy cells do
not directly interact with ill cells, so $R_{21}=0$. The carrying capacity
limitation for ill cells, while it may exist, is considered to be much 
larger than that for healthy cells, which allows us to set $R_{22}=0$. 
Ill cells are killed and eliminated by immune cells, hence $R_{23}
\equiv-A_{23}$ and $R_{24} \equiv-A_{24}$. The degradation of ill 
cells is increased under the influence of pathogens as the latter 
catalyze the immune system, hence $R_{25}\equiv-A_{25}$. The number 
of ill cells rises as a result of pathogens infecting healthy cells, 
which gives $F_2=A_{51}N_5N_1$.

\vskip 2mm
(iii) {\it Innate immune cells} die by apoptosis, with a rate $R_3\equiv-
A_3$. They can be promoted by healthy cells, so $R_{31}\equiv A_{31}$.
And they are activated by ill cells, $R_{32}\equiv A_{32}$. The carrying
capacity of immune cells is much larger than that of healthy cells, which
makes it admissible to set $R_{33}=0$. Innate immune cells are activated
by specific immune cells, hence $R_{34}\equiv A_{34}$. And they are
activated by pathogens, $R_{35}\equiv A_{35}$. There is no external
flux from outside of the organism, that is $F_3=0$.

\vskip 2mm
(iv) {\it Specific immune cells} also have a finite lifetime, characterized
by an apoptosis rate $R_4\equiv-A_4$. They are promoted by healthy cells,
$R_{41}\equiv A_{41}$, and are activated by ill cells, $R_{42}\equiv A_{42}$.
Innate immune cells inhibit an excessive amount of specific immune cells,
$R_{43}\equiv-A_{43}$. Similarly to innate cells, for specific immune
cells, the carrying capacity limitation can be ignored, so that $R_{44}=0$.
Pathogens activate specific immune cells, hence $R_{45}\equiv A_{45}$. And
there is no external flux, $F_4=0$.

\vskip 2mm
(v) {\it Pathogens} are characterized by a natural decay rate $R_5\equiv-
A_5$. Their number does not depend on the number of healthy cells, $R_{51}=0$.
Pathogens proliferate by the lysis of ill cells, $R_{52}\equiv A_{52}$. They
are killed and eliminated by innate, as well as specific immune cells, hence
$R_{53}\equiv-A_{53}$ and $R_{54}\equiv-A_{54}$. The number of pathogens can
be  an order of or several orders larger than that of healthy cells. Therefore,
there is practically no carrying capacity limitation for them, that is,
it is safe to set $R_{55}=0$. Contrary to all other organism cells, pathogens
are supplied from the external surrounding, therefore $F_5=F$ is not zero,
but is to be treated as a parameter characterizing the environment in which the
organism lives. In this letter, we only focus on a constant pathogen flux.

Taking into consideration the described processes transforms Eq. (\ref{eq1})
to the set of five evolution equations
$$
\frac{dN_1}{dt} = A_1N_1 - A_{11}N_1^2 - A_{13}N_1N_3 -
A_{14}N_1N_4 - A_{15}N_1N_5  \; ,
$$
$$
\frac{dN_2}{dt} = - A_2N_2 - A_{23}N_2N_3 -
A_{24}N_2N_4 - A_{25}N_2N_5 + A_{51}N_5N_1 \; ,
$$
$$
\frac{dN_3}{dt} = - A_3N_3 + A_{31}N_3N_1 +
A_{32}N_3N_2 + A_{34}N_3N_4 + A_{35}N_3N_5  \; ,
$$
$$
\frac{dN_4}{dt} = - A_4N_4  + A_{41}N_4N_1 +
A_{42}N_4N_2 - A_{43}N_4N_3 + A_{45}N_4N_5  \; ,
$$
\be
\label{eq4}
\frac{dN_5}{dt} = - A_5N_5 + A_{52}N_5N_2 -
A_{53}N_5N_3 - A_{54}N_5N_4  + F \; .
\ee

Using explicitly Eqs. (\ref{eq4}) is not convenient, since the numbers
of cells are extremely large. For instance, the number of healthy cells is
$N_1\sim 10^{13}$, and the number of pathogens can be as large as $N_5\sim
10^{14}$ or more. Therefore, to be practical, Eqs. (\ref{eq4}) are to be normalized
by introducing a normalization constant $N$ and defining the cell fractions
\be
\label{eq5}
x_i \equiv \frac{N_i}{N} \qquad (i=1,2,3,4,5) \; .
\ee
Note that the cell fractions do not sum up to $1$, since $N$ is arbitrary 
(but is introduced so that we deal with quantities of order $1$).

Also, it is necessary to determine a time scale characterizing a typical
duration of the homeostasis processes. We thus introduce a typical time 
scale $\tau$. Then we define the dimensionless decays rates
\be
\label{eq6}
\al_i \equiv A_i \tau
\ee
and the dimensionless interaction parameters
\be
\label{eq7}
a_{ij} \equiv A_{ij} N \tau \; .
\ee
In addition, we define the dimensionless pathogen influx
\be
\label{eq8}
\vp \equiv \frac{\tau}{N}\; F \; .
\ee

Making use of the dimensionless quantities, and measuring time in units of
$\tau$, we reduce Eqs. (\ref{eq4}) to the dynamical system
\be
\label{eq9}
\frac{dx_i}{d\tau} = f_i  \qquad (i=1,2,3,4,5) \; ,
\ee
with the right-hand sides
$$
f_1 = \al_1 x_1 - a_{11}x_1^2 - a_{13}x_1 x_3 - a_{14}x_1 x_4 -
a_{15}x_1 x_5  \; ,
$$
$$
f_2 = - \al_2 x_2 - a_{23}x_2 x_3 - a_{24}x_2 x_4 -
a_{25}x_2 x_5 + a_{51} x_5 x_1\; ,
$$
$$
f_3 = - \al_3 x_3 + a_{31}x_3 x_1 + a_{32}x_3 x_2 +
a_{34}x_3 x_4 + a_{35}x_3 x_5  \; ,
$$
$$
f_4 = - \al_4 x_4 + a_{41}x_4 x_1 + a_{42}x_4 x_2 -
a_{43}x_4 x_3 + a_{45}x_4 x_5  \; ,
$$
\be
\label{eq10}
f_5 = - \al_5 x_5 + a_{52}x_5 x_2 - a_{53}x_5 x_3 -
a_{54}x_5 x_4 + \vp \; .
\ee
Equations (\ref{eq9}) and (\ref{eq10}) are the basic equations describing
the organism homeostasis.

\section{Structural stability analysis}

The dynamical system, given by Eqs. (\ref{eq9}) and (\ref{eq10}), contains
25 parameters $\al_j$ and $a_{ij}$, which appears to be an insuperable obstacle
for studying its properties. The situation can be simplified by choosing
appropriate scaling parameters $N$ and $\tau$. Thus, for the normalization
constant $N$, we can take the capacity number of healthy cells
\be
\label{eq11}
N = \frac{A_1}{A_{11}} \; .
\ee
As the temporal scale $\tau$, it is reasonable to choose the characteristic
time of healthy-cell reproduction
\be
\label{eq12}
\tau = \frac{1}{A_1} \; .
\ee
With these scaling parameters, we have
\be
\label{eq13}
\al_1 = 1 \; , \qquad a_{11} =  1 \; .
\ee
These values characterize the typical rates and interactions involved in
the homeostasis of the organism.

As is noted above, ill cells can either exhibit a natural decay, when
$\al_2>0$, or can show a pathological proliferation, when $\al_2<0$. It 
is convenient to use the notation
\be
\label{eq14}
\bt \equiv \frac{1-\al_2}{2} \; , \qquad \al_2 =  1 - 2\bt \; .
\ee
The value $\beta = 0$ corresponds to $\al_2=1 >0$, while $\beta = 1$ 
corresponds to $\al_2=-1 <0$.

We may assume that the innate and specific immune cells enjoy the same
apoptosis rate
\be
\label{eq15}
\al \equiv \al_3 =  \al_4 \; .
\ee
And let us set $\al_5=1$. We denote the parameters, associated with the
interactions between immune and healthy cells as
\be
\label{eq16}
b \equiv a_{13} = a_{31} = a_{14} = a_{41} \; .
\ee
When $b=0$, immune cells do not attack healthy cells, hence there are 
no autoimmune diseases. Conversely, for $b>0$, autoimmune disorders 
become possible. We also may assume that all other processes, except 
those related to Eq. (\ref{eq16}), are characterized by the intensities 
being comparable to $a_{11}$ (equal to $1$ according to Eq. (\ref{eq13})). 
That is, we can put
\be
\label{eq17}
a_{ij} = 1 \qquad (a_{ij} \neq b) \; .
\ee
In this way, we are left with four parameters, $\al$, $\bt$, $b$, and $\vp$.

The values of the parameters $\bt$ and $b$ control the occurrence of a 
chronic pathology or of an autoimmune disorder. Varying these parameters, 
we can reach four limiting cases.

\vskip 2mm
(i) {\it No chronic pathology and no autoimmune disorder}:
\be
\label{eq18}
\bt = 0 \; , \qquad  b = 0 \; .
\ee

\vskip 2mm
(ii) {\it No chronic pathology with autoimmune disorder}:
\be
\label{eq19}
\bt = 0 \; , \qquad  b = 1 \; .
\ee

\vskip 2mm
(iii) {\it Chronic pathology but no autoimmune disorder}:
\be
\label{eq20}
\bt = 1 \; , \qquad  b = 0 \; .
\ee

\vskip 2mm
(iv) {\it Chronic pathology and autoimmune disorder}:
\be
\label{eq21}
\bt = 1 \; , \qquad  b = 1 \; .
\ee

We can change the parameters $\bt$ and $b$ in the range between these
limiting cases, thus, considering a variety of different organisms. For
each fixed pair of $\bt$ and $b$, we vary the parameters $\al$ and $\vp$
in a wide range, finding all admissible fixed points of the dynamical
system. In the standard way, we accomplish the Lyapunov stability analysis
and select the stable fixed points. In order to present the results of our
analysis in the most illuminating form, it is useful to define the sum
of the healthy-cell fraction and of the ill-cell fraction,
\be
\label{eq22}
x \equiv x_1 + x_2 \; .
\ee
Also, we introduce the summary fraction of all immune cells
\be
\label{eq23}
y \equiv x_3 + x_4 \; .
\ee

The main, and to some extent surprising, conclusion of the stability 
analysis for the dynamical system (\ref{eq9}) is its remarkable structural 
stability. Varying the parameters $\bt$ and $b$ between the qualitatively
different cases (\ref{eq18}) to (\ref{eq21}) always results, for any given
$\bt$ and $b$, in four stationary organism states characterized by the
fixed-point values $x^*$ and $y^*$.

\vskip 2mm

{\bf $\bf A$. Alive state}:

\vskip 2mm

\be
\label{eq24}
x^* > 0 \; , \qquad y^* > 0 \; ,
\ee
when there are both self-cells and immune cells.

\vskip 2mm

{\bf $\bf B$. Boundary state}:

\vskip 2mm

\be
\label{eq25}
x^* > 0 \; , \qquad y^* = 0 \; ,
\ee
when there are self-cells, but no immune cells.

\vskip 2mm

{\bf $\bf C$. Critical state}:

\vskip 2mm

\be
\label{eq26}
x^* = 0 \; , \qquad y^* > 0 \; ,
\ee
when only immune cells survive.

\vskip 2mm

{\bf $\bf D$. Dead state}:

\vskip 2mm

\be
\label{eq27}
x^* = 0 \; , \qquad y^* = 0 \; ,
\ee
when there are neither self-cells nor immune cells.

\vskip 2mm

All these states are always present in the phase diagram on the $\al-\vp$
plane. The boundaries between the states, of course, are different for
different $\bt$ and $b$, but all four states do remain stable.

The exact domains of stability for each of the states (\ref{eq24}) to
(\ref{eq27}), as well as the values of all stationary fractions $x_j^*$,
have been found numerically and, in some cases, analytically. Here, we 
illustrate the results by the phase portraits corresponding to the 
limiting cases (\ref{eq18}) to
(\ref{eq21}).

The case (\ref{eq18}) of no chronic pathology and no autoimmune disorder
is represented in the phase portrait of Fig 1. The case (\ref{eq19}) of
no chronic pathology but with autoimmune disorder is shown in Fig 2. In
Fig. 3, the case (\ref{eq20}) is demonstrated, when there is chronic
pathology but there is no autoimmune disorder. And Fig. 4 illustrates
the case (\ref{eq21}), when there exist both chronic pathology as well
as autoimmune disorder. As is seen, the boundaries between the states
move when varying the system parameters, but all four states (\ref{eq24})
to (\ref{eq27}) are always present. The case (\ref{eq20}) differs from
other cases by the presence of a narrow region of bistable states. In 
this region, there exist large fluctuations of the coexisting states 
[8,9]. The dead state (\ref{eq27}) is always located in the same part 
of the phase diagram.

We have also studied the case, when the pathogen influx randomly 
fluctuates around its mean value (\ref{eq8}). In that case, the cell 
fractions also fluctuate in time. Then the picture, presented here, 
corresponds to the behavior of guiding centers that are defined by 
the averaging method [10].

Concluding, we have constructed a dynamical system describing the 
homeostasis of an organism as a whole. The organism is composed of five
types of components: healthy cells, ill cells, innate and specific 
immune cells, and pathogens. A principal novelty in formulating the 
model is that we consider not some parts of an organism, but treat it as 
a total self-consistent system subject to the influence of pathogens. 
Another important point in constructing the evolution equations is that 
we take into account the action-counteraction symmetry (\ref{eq3}). By 
varying the system parameters in a very wide range, we have shown that 
the dynamical system (\ref{eq9}) enjoys a remarkable structural stability, 
always exhibiting four stable stationary states. These results remain 
valid if we include in the dynamical system (\ref{eq9}) nonzero values of 
the carrying capacity limitations for ill cells and immune cells. We have 
also accomplished direct numerical solution of Eqs. (\ref{eq9}) and 
(\ref{eq10}), confirming the structural stability of the dynamical system. 

Our basic aim here has been to present the new model of a biological 
organism and to accomplish a detailed stability analysis making it 
possible to find numerically the typical phase portraits. An important 
result of the present paper is the remarkable structural stability of the 
studied biological system. The found four stable states characterize all 
basic states of an organism: alive relatively healthy state, boundary ill 
state, critically ill state, and the dead state. 

We do not overload this paper by discussing various admissible medical  
interpretations and applications. This analysis will be the topic of 
separate publications.

\vskip 5mm

{\bf Acknowledgement}. We are grateful for financial support to the 
Competence Center for Cooping with Crisis in Socio-Economic Systems at 
ETH Zurich. One of the authors (V.I.Y.) acknowledges helpful 
discussions with E. Kapuscik.

\newpage

\begin{center}

{\bf{\Large Figure captions} }

\end{center}

\vskip 1cm

{\bf Fig. 1}. Phase portrait on the $\al-\vp$ plane for the case
(\ref{eq18}) of no chronic pathology ($\bt=0$) and no autoimmune
disorder ($b=0$), showing the stability domains of the stationary
states (\ref{eq24}) to (\ref{eq27}).

\vskip 1cm

{\bf Fig. 2}. Phase portrait for the case (\ref{eq19}) of no
chronic pathology ($\bt=0$) but with autoimmune disorder ($b=1$),
demonstrating the stability regions of the stationary states (\ref{eq24})
to (\ref{eq27}).

\vskip 1cm

{\bf Fig. 3}. Phase portrait for the case (\ref{eq20}), when there exists
chronic pathology ($\bt=1$) but there is no autoimmune disorder ($b=0$).
The peculiarity here is in the occurrence of bistable states.

\vskip 1cm

{\bf Fig. 4}. Phase portrait for the case (\ref{eq21}), when both
chronic pathology ($\bt=1$) and autoimmune disorder ($b=1$) are present.
The classification of the stationary states is the same as in Eqs.
(\ref{eq24}) to (\ref{eq27}).

\newpage

\begin{figure}[ht]
\center
\includegraphics[width=12cm]{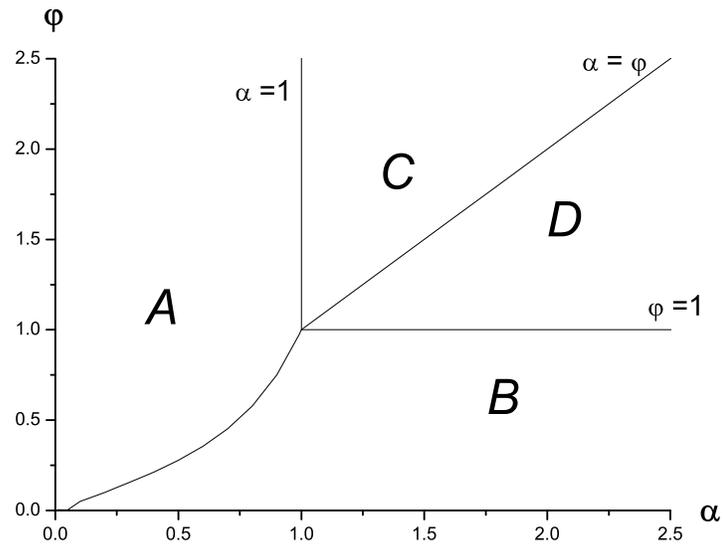}
\caption{Phase portrait on the $\al-\vp$ plane for the case (\ref{eq18})
of no chronic pathology ($\bt=0$) and no autoimmune disorder ($b=0$),
showing the stability domains of the stationary states (\ref{eq24})
to (\ref{eq27}).}
\label{fig:Fig.1}
\end{figure}

\begin{figure}[ht]
\centerline{\includegraphics[width=12cm]{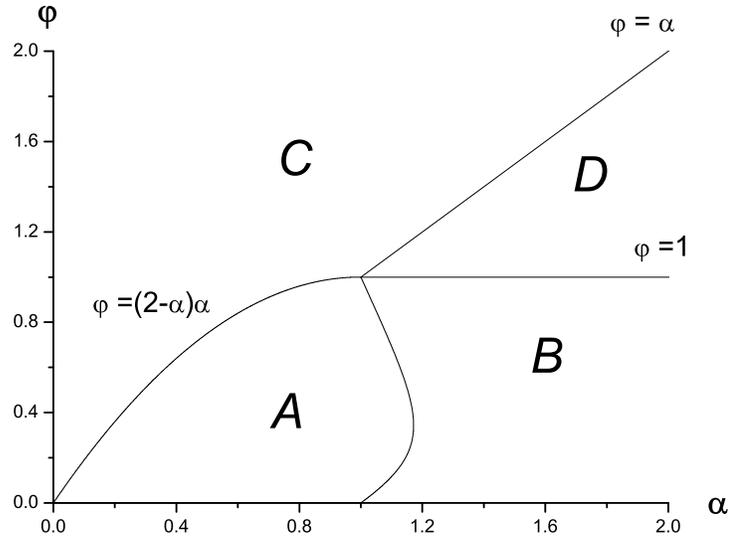}}
\caption{Phase portrait for the case (\ref{eq19}) of no chronic pathology
($\bt=0$) but with autoimmune disorder ($b=1$), demonstrating the stability
regions of the stationary states (\ref{eq24}) to (\ref{eq27}).}
\label{fig:Fig.2}
\end{figure}

\begin{figure}[ht]
\centerline{\includegraphics[width=12cm]{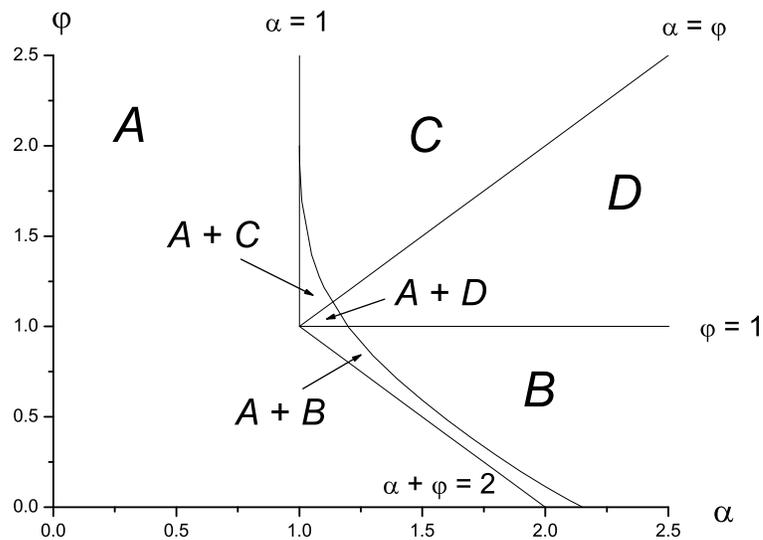}}
\caption{Phase portrait for the case (\ref{eq20}), when there exists
chronic pathology ($\bt=1$) but there is no autoimmune disorder ($b=0$).
The peculiarity here is in the occurrence of bistable states.}
\label{fig:Fig.3}
\end{figure}

\begin{figure}[ht]
\centerline{\includegraphics[width=12cm]{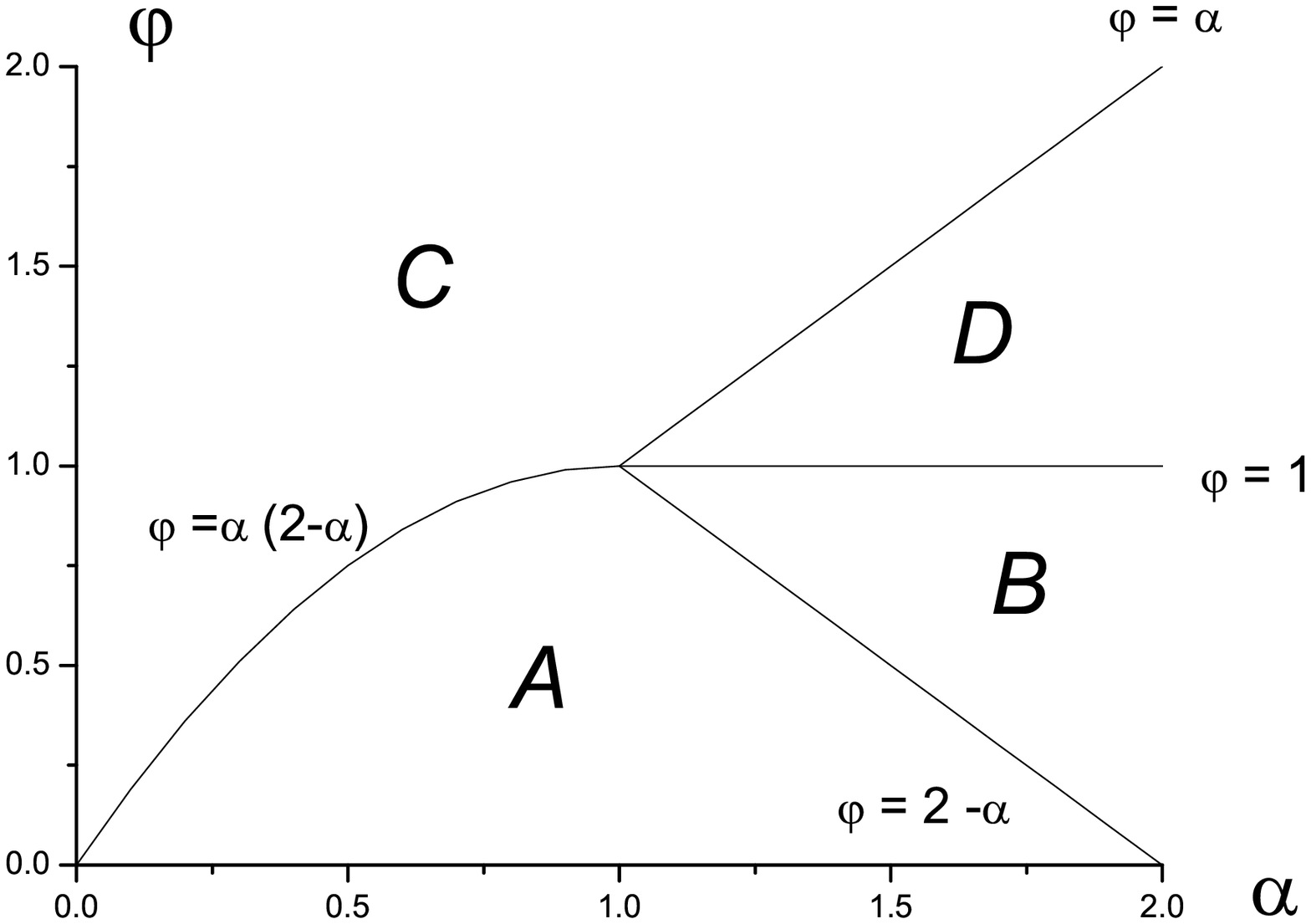}}
\caption{Phase portrait for the case (\ref{eq21}), when both chronic
pathology ($\bt=1$) and autoimmune disorder ($b=1$) are present. The
classification of the stationary states is the same as in Eqs. (\ref{eq24})
to (\ref{eq27}).}
\label{fig:Fig.4}
\end{figure}


\newpage

\begin{thebibliography}{99}
\bibitem{1}
I. Prigogine, {\it From Being to Becoming} (Freeman, 
San Francisco, 1980).

\bibitem{2}
G. Nicolis and I. Prigogine, {\it Exploring Complexity} (Freeman, 
New York, 1989).

\bibitem{3}
A.S. Perelson and G. Weisbuch, {\it Rev. Mod. Phys.} {\bf 69} (1997) 
1219.

\bibitem{4}
R. Thom, {\it Structural Stability and Morphogenesis} (Benjamin, 
Reading, 1975).

\bibitem{5}
A.J. Lichtenberg and M.A. Lieberman, {\it Regular and Stochastic Motion}
(Springer, New York, 1963).

\bibitem{6} 
D. Sornette, A. B. Davis, K. Ide, K. R. Vixie, V. Pisarenko and J. R. Kamm,
{\it Proc. Nat. Acad. Sci. USA} {\bf 104} (2007) 6562.

\bibitem{7}
D. Sornette, V.I. Yukalov, E.P. Yukalova, J.Y. Henry, D. Schwab
and J.P. Cobb, arXiv:0710.3859 (2007).

\bibitem{8}
V.I. Yukalov, {\it Phys. Rep.} {\bf 208} (1991) 395.

\bibitem{9}
V.I. Yukalov, {\it Int. J. Mod. Phys. B} {\bf 17} (2003) 2333.

\bibitem{10}
N.N. Bogolubov and Y.A. Mitropolsky, {\it Asymptotic Methods in the 
Theory of Nonlinear Oscillations} (Gordon and Breach, New York, 1961).

\end{thebibliography}
\end{document}